# An Ensemble Deep Learning Model for Drug Abuse Detection in Sparse Twitter-Sphere

Han Hu[a], NhatHai Phan[a], James Geller[a], Stephen Iezzi[a], Huy Vo[b], Dejing Dou[c], Soon Ae Chun[d]

[a] *Ying Wu College of Computing, New Jersey Institute of Technology, Newark, NJ, USA,*
[b] *Department of Computer Science, The City College of New York, New York, NY, USA,*
[c] *Computer and Information Science, University of Oregon, Eugene, OR, USA*
[d] *Information Systems & Informatics, City University of New York, Staten Island, NY, USA*

**Abstract**

*As the problem of drug abuse intensifies in the U.S., many studies that primarily utilize social media data, such as postings on Twitter, to study drug abuse-related activities use machine learning as a powerful tool for text classification and filtering. However, given the wide range of topics of Twitter users, tweets related to drug abuse are rare in most of the datasets. This imbalanced data remains a major issue in building effective tweet classifiers, and is especially obvious for studies that include abuse-related slang terms. In this study, we approach this problem by designing an ensemble deep learning model that leverages both word-level and character-level features to classify abuse-related tweets. Experiments are reported on a Twitter dataset, where we can configure the percentages of the two classes (abuse vs. non abuse) to simulate the data imbalance with different amplitudes. Results show that our ensemble deep learning models exhibit better performance than ensembles of traditional machine learning models, especially on heavily imbalanced datasets.*

***Keywords:*** *Substance-Related Disorders, Machine Learning, Social Media.*

## Introduction

Misuse and abuse of prescription drugs and of illicit drugs have been major public health problems in the United States for decades. A "Public Health Emergency" declared in 2016 [1] and several official surveys [2] all show that the problem has been getting worse in recent years. For example, the most recent reports from the National Survey on Drug Use and Health (NSDUH) [2] estimate that 10.6% of the total population of people ages 12 years and older (i.e., about 28.6 million people) have misused illicit drugs in 2016, which represents an increase of 0.5% over 2015. According to the Centers for Disease Control and Prevention (CDC), opioid drugs were involved in 42,249 known deaths in 2016 nationwide [3]. In addition, the number of heroin-related deaths has been increasing sharply over five years and has surpassed the number of firearm homicides in 2015 [4]. The emerging new problems, such as the epidemic of illicitly manufactured fentanyl (IMF) [5], marijuana-related traffic accidents [6], and marijuana use among adolescents [7] are posing further increasing threats to public health.

To fight this epidemic of drug abuse, methods of social media monitoring with wider scope and shorter response time are needed. Social media, such as Twitter, have been proven to be sufficient and reasonably reliable data sources for social-level detection and monitoring tasks [8]. Twitter is a popular social media platform that has 100 million daily active users and 500 million daily tweets [10] (messages posted by Twitter users), most of which are publicly accessible, on a wide range of topics.

We are using algorithms for filtering and classification for acquiring abuse-related tweets for analysis and monitoring. Filtering is the very first and most basic step toward extracting potentially useful tweets from the large number posted every day. Filtering, by itself, even with standard drug names (e.g. heroin), generally does not suffice to produce a dataset pure enough for practical use. Thus, machine learning classifiers have to be trained to further identify tweets that are related to drug abuse. However, most abuse-related Twitter datasets have the problem of imbalanced class distributions. Typical datasets, collected with only the names of drugs, may have 5% to 30% of positive (abuse-related) tweets, due to the topic diversity and language irregularity of tweets. The percentage of positive tweets decreases sharply when more keywords, especially slang names for drugs (e.g., snow) and abuse behavior keywords (e.g., snorting), are included in a tweet dataset. The imbalanced class distribution and the noisy nature of the Twitter data make it hard to train a classifier with good performance.

In this paper, we propose an ensemble of two types of deep learning-based methods as better options, among classifiers, for situations in which the collected data is inevitably imbalanced, because they are more robust than traditional machine learning models. Our ensemble deep learning model combines word-level CNN models and character-level CNN models to perform classification. We compare our models with baseline models on a dataset we collected, where we can configure the class distribution of positive versus negative tweets in the training data and test data. By changing the percentage of positively and negatively labeled data in the dataset, we can simulate the imbalanced datasets that were collected by different means. We validate the performance of different models in a variety of settings to get a clearer picture of how imbalanced data affect classification performance.

## Related Works

Large scale surveys, such as NSDUH [2], Monitoring the Future [11], the MedWatch program [12], and the results derived from these surveys [13], clearly show that there is an epidemic of drug abuse across the United States. However, a recent report [14] states that the estimated number of deaths due to prescription drugs could be inflated due to the difficulties in determining whether a drug is obtained by prescription or not. We assert that the ambiguities highlighted in this new report raise questions about the reliability of the earlier surveys, and thus, such a report illustrates the potential value of social media-based studies.

In fact, several studies found positive correlations between Twitter data and real world data. Chary et al. [15] performed semantic analysis on 3.6 million tweets with 5% labeled and found significant agreement with the NSDUH data. Hanson et al. [16] conducted a quantitative analysis on 213,633 tweets discussing Adderall, and found positive geo-temporal correlations. Furthermore, Shutler et al. [17] performed a qualitative analysis of prescription opioid-related tweets and found that indications of abuse were common. On the other hand, several studies focused on designing machine learning models to preform tweet classification. Mahata et al. [18] performed a comprehensive study on using deep learning models to identify mentions of drug intake in tweets. Katsuki et al. [20] trained SVM on a dataset of 1,000 tweets for classification of tweets for relevance and favorability of online drug sales. Hu et al. [19] showed the potential of applying deep learning models in a drug abuse monitoring system to detect abuse-related tweets. Sarker et al. [9] proposed an ensemble of traditional machine learning models to classify drug abuse tweets and non-abuse tweets of certain drugs. Other studies focus on social media users, such as Fan's work [26] utilizes user interaction networks to identify opioid users on Twitter. In this paper, we will be developing ensemble deep learning models to expand the classification of tweets to a border scope of drugs and their abuse behaviors with better performance in the unbalanced class distribution settings.

## Methods

In this section, we present the definition of the *drug abuse-related risk behavior detection problem*, our methods for collecting tweets, our methods for labeling tweets, and our ensemble deep learning approach.

### Problem Definition

In this paper, our first goal is to build a Twitter dataset consisting of tweets that are related to drug abuse risk behaviors (**positive** tweets), and tweets that are not (**negative** tweets). The "drugs" in the term "drug abuse risk behaviors" in this study include Schedule 1 and Schedule 2 drugs and their derivatives [21], including marijuana, heroin, cocaine, fentanyl, etc. The reasons that we include *marijuana* even though it is legalized in several states are that: **(1)** Marijuana is still a controlled substance in the federal law, whether for medical use or recreational use; and **(2)** Marijuana can still cause harm to adolescents [7], can cause "use disorder" [13], and is related to traffic fatalities [6]. The term "abuse risk behavior" can be defined as "The existence of likely abusive activities, consequences, and endorsements of drugs." Tweets that contain links to or summarize news and reports related to drug abuse, and tweets that merely express opinions about drug abuse, are counted as negative in this study. Our main goal in this paper is to train a model that can accurately classify positive and negative tweets in a highly imbalanced (drug abuse) dataset.

### Data Collection

Although there are human-labeled drug abuse Twitter datasets (e.g. Sarker's dataset [9]) available, due to Twitter's data policy, which prohibits the direct sharing of tweet contents, by the time we access the tweets in that dataset, more than 40% of tweets are either removed or hidden from the public. This significantly affects the quality and integrity of existing publicly available datasets. Therefore, we need to build a new dataset from scratch. In our framework, raw tweets are collected through a set of Application Programming Interfaces (Twitter APIs) via keyword filtering. By defining a set of keywords, the API will fetch tweets that contain any of the keywords from either the real-time stream of tweets or from archived tweets. For a more complete coverage of drug-related topics, we selected three types of keywords: **(1)** Full and official names of drugs, e.g. marijuana, cocaine, OxyContin, fentanyl, etc.; **(2)** Slang terms for drugs, e.g. pot, blunt, coke, crack, smack, etc.; and **(3)** Drug abuse-related behaviors and symptoms, e.g., high, amped, addicted, headache, dizzy, etc. The number of keywords we used is limited to 400 by the Twitter APIs.

### Data Annotation

We build a comperhaensive guide, accessable at **https://goo.gl/tqWddS**, based on Sarker's guide [9] . Each one of the three members in out research team with experience in health informatics annotates the 1,794 tweets from Hu et al.'s study [19] independently following the guide. A final label for each seed tweet is determined by majority voting from three labels.

To acquire annotated tweets rapidly, at low cost, and with increased percentage of positive tweets, we **(1)** use these labeled tweets as *"seed"* tweets to train a SVM classifier; **(2)** run the SVM classifier on the unlabeled dataset, and randomly sample 5,000 machine labeled tweets that have prediction probability (esitmated with Platt scaling) > 0.8; and **(3)** post the 5,000 tweets (without identification information) onto the Amazon Mechanical Turk (AMT) crowdsourcing platform for annotation. AMT is a well-known crowdsourcing platform where Posters can post Human Intelligence Tasks (HITs) and Workers finish HITs for micro-payments. A literature study [22] evaluated AMT as a thrustworthy platform to obtain human labeled data. The same guide is used to guide the Workers how to annotate the tweets. Each tweet is posted as one HIT that requires the Worker to label it as positive or negative following the guide. Each HIT is replicated as three *assignments* to be completed by three individual Workers. We set the price of each assignment to be $0.05, a very generous price compared to what was reported in Buhrmester's work [22] All HITs are completed within hours after being posted. The final label of each tweet is aggregated from the three labels by majority voting. We also label 1,000 tweets randomly sampled from the 5,000 tweets with our annotator as a measure of quality check.

*Table 1-Details of Pre-trained Word Embedding*

| Name | Model | Corpus | Dimension |
| --- | --- | --- | --- |
| GoogleNews | Word2vec | ~100 billion words | 300 |
| Glove Common | Glove | ~42 billion words | 300 |
| Godin | Word2vec | ~400 million tweets | 400 |
| Drug Chatter | Word2vec | ~1 billion tweets | 400 |

### Feature Extraction

Machine learning models require numerical features to work with. Feature extraction transforms text features into numerical features in the form of vectors. To cover the content ambiguity in drug abuse-related tweets, a variety of feature extraction methods are used in this study. In our word-level CNN models, we use pre-trained word embedding models that were trained on large corpora to transform words into dense vectors. We test several pre-trained models as Mahata's work [18] suggested. With our word-level CNN model, the *Drug Chatter* embedding has the best average performance on our dataset; thus, it is chosen as the pre-trained word embedding model for this study. The details of the tested word embedding models are shown in Table 1. Each tweet is converted to a sequence of 400-dimensional vectors. Considering that the length limit of each tweet nowadays is 280 chars, the sequence length is set to 40. In our char-level CNN, the preprocessing step only turns all characters to lower case as suggested by [23] . Each char is then converted into a 128-dimensional trainable randomly-initialized vector. Instead of being fixed, the character embeddings are trained along with other layers in the model.

We also replicate the features extracted in Sarker et al. study [9] , including: **(1)** The tokenization process; **(2)** The abuse-indicating term features, consisting of the presence and the counts of abuse-indicating terms obtained from Hanson et al. [16] ; **(3)** The drug-slang lexicon features, consisting of the presence and the counts of terms longer than five characters found in an online drug abuse dictionary [24] ; **(4)** The word cluster features, represented by 150-dimensional one-hot vectors, were constructed by identifying words that belong to certain word clusters in a dataset [9] that contains 150 drug-related word clusters; and **(5)** The synonym expansion features, accomplished by identifying all synonyms of all nouns, verbs, and adjectives in the tokenized tweets using WordNet [25] .

*An Ensemble Deep Learning Model for Drug Abuse Detection in Sparse Twitter-Sphere*

In this section, we present our novel ensemble deep learning model for drug abuse risk behavior detection by integrating extracted features from tweets into CNN models. Our ensemble model takes the outputs of multiple prediction models, word-level CNN (**W-CNN**) and char-level CNN (**C-CNN**) [29] in our case, and feed them to a meta-learner that gives the final predictions. We design W-CNN and C-CNN for this task. In fact, both the W-CNN and the C-CNN share a similar structure as shown in Figure 1.

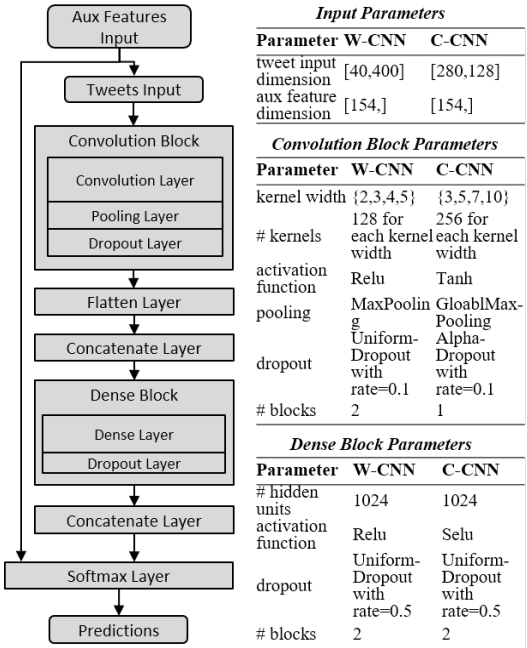

*Figure 1- Ensemble CNN model structures.*

The inputs of our W-CNN are vectors of shape [40, 400] where 40 is the maximum sequence length (number of words allowed) in an input tweet, and 400 is the length of the pre-trained word embeddings. The input of our C-CNN is shaped as [280, 128] where 280 is the maximum possible length of a tweet, and 128 is the length of the vector representation of each character in the charset. The auxiliary features in the input include: **(1)** The synonym expansion features in the form of synonymous words are directly concatenated with the input tweets (before they are transformed into vectors); and **(2)** The remaining auxiliary features, in the form of 154-dimensional vectors, are concatenated to the last hidden layer of the dense layers. For each convolution kernel size, the W-CNN model has two convolution layers with ReLU activation functions stacked together. Each is followed by a max-pooling layer.

The C-CNN model has one convolution layer for each convolution kernel size with Tanh activation function, followed by a global-max-pooling layer that performs max-pooling over all outputs of convolution layers with different kernel sizes. Both models have one dense layer block, consisting of two dense layers with 1,024 hidden units each, and one Softmax output layer with two units. The activation functions are slightly different, as the W-CNN model uses ReLU, while the C-CNN model uses SELU. The output of the last hidden layer is concatenated with vectors of abuse-indicating term features, drug-slang lexicon features, and word cluster features, before being fed into the output layer.

Finally, a number of independently trained CNN models of both types are ensembled together by using majority. Model ensembles were also used in Sarker et al. study [9] to reduce variability and bias, in order to improve prediction performance. We apply the same ensemble strategy to both our deep learning models and the baseline models.

*Experimental Design*

Our main objective in this experiment is to directly compare the performances of the ensemble traditional machine learning model and the ensemble deep learning model. For the ensemble traditional machine learning model, two of each type of baseline models, six in total, are trained and ensembled together. For the ensemble deep learning model, six models of three types (two for each type) are used. The three types are denoted as follows. **(1)** "*char_aux*" is the char-level CNN model with auxiliary features. **(2)** "*char_cnn*" is the plain char-level CNN without any auxiliary features. **(3)** "*word_aux*" is the word-level CNN model with all auxiliary features. For deep learning models, it is extremely easy to overfit, due to the rather small number of training and test data elements; thus, the model is saved at each training epoch, and the best epoch is found among the saved models. For each class distribution scenario, each model is trained with the same six sets of training data and tested on the corresponding test data. All results reported are averaged results from the 6-fold cross-validation.

*Table 2. Dataset Variants*

| Class distribution (positive: negative) | # of training data items | # of test data items |
|---|---|---|
| 50:50 split | 3450 | 690 |
| 40:60 split | 2850 | 570 |
| 30:70 split | 2450 | 490 |
| 20:80 split | 2150 | 430 |
| 10:90 split | 1900 | 380 |

## Experimental Results

*Data Annotation Results*

From Jan 2017 to Feb 2017, we collected 3,265,153 tweets in total. The *"seed"* dataset that we annotated to be used to train the pre-filter consistes of 1,794 tweets, including 280 positive labels and 1,514 negative labels. Our annotator achieved the agreement score of 0.414, measured by Krippendoff's Alpha. For the AMT labeled dataset, we removed duplicate tweets from it, resulting dataset contains 4,736 tweets with 2,657 positive labels and 2,079 negative labels. The agreement score is 0.456 measured by Alpha, which can be considered as a reliable result in our study as demostrated in [27] , since: **(1)** We are performing data annotation with data aggregation to reduce variability, instead of typical content analysis [28]; **(2)**

*Table 3-Experimental Results*

| | Class Distribution: 50:50 split | | | | | | | |
|---|---|---|---|---|---|---|---|---|
| Measure | Ensemble CNN | Ensemble ML | char_aux | char_cnn | word_aux | SVM | Random Forest | Naive Bayes |
| Accuracy | 0.8510 | **0.8575** | 0.8506 | 0.8477 | 0.8466 | 0.8415 | **0.8586** | 0.8384 |
| Precision_p | **0.8468** | 0.8350 | 0.8315 | 0.8240 | 0.8198 | 0.8063 | **0.8404** | 0.8319 |
| Recall_p | 0.8575 | **0.8918** | 0.8797 | 0.8845 | 0.8894 | **0.9000** | 0.8860 | 0.8493 |
| F1 score_p | 0.8520 | **0.8623** | 0.8549 | 0.8531 | 0.8529 | 0.8504 | **0.8624** | 0.8402 |
| | Class Distribution: 40:60 split | | | | | | | |
| Measure | Ensemble CNN | Ensemble ML | char_aux | char_cnn | word_aux | SVM | Random Forest | Naive Bayes |
| Accuracy | 0.8567 | **0.8582** | 0.8528 | **0.8563** | 0.8430 | 0.8444 | 0.8494 | 0.8427 |
| Precision_p | **0.8079** | 0.8047 | 0.8007 | 0.8055 | 0.7818 | **0.8104** | 0.7770 | 0.7862 |
| Recall_p | 0.8428 | **0.8531** | 0.8421 | 0.8454 | 0.8443 | 0.7982 | **0.8746** | 0.8341 |
| F1 score_p | 0.8249 | **0.8280** | 0.8207 | **0.8248** | 0.8113 | 0.8041 | 0.8229 | 0.8093 |
| | Class Distribution: 30:70 split | | | | | | | |
| Measure | Ensemble CNN | Ensemble ML | char_aux | char_cnn | word_aux | SVM | Random Forest | Naive Bayes |
| Accuracy | **0.8599** | 0.8595 | 0.8522 | 0.8507 | 0.8483 | 0.8429 | **0.8537** | 0.8452 |
| Precision_p | 0.7402 | **0.7426** | 0.7253 | **0.8223** | 0.718 | 0.7467 | 0.7137 | 0.7218 |
| Recall_p | 0.8231 | 0.8163 | 0.8209 | 0.8180 | 0.8158 | 0.7234 | **0.8583** | 0.7914 |
| F1 score_p | **0.7792** | 0.7771 | 0.7695 | 0.7666 | 0.7635 | 0.7336 | **0.7789** | 0.7538 |
| | Class Distribution: 20:80 split | | | | | | | |
| Measure | Ensemble CNN | Ensemble ML | char_aux | char_cnn | word_aux | SVM | Random Forest | Naive Bayes |
| Accuracy | **0.8674** | 0.8508 | **0.8624** | 0.8568 | 0.8506 | 0.8384 | 0.8475 | 0.8527 |
| Precision_p | **0.6416** | 0.5908 | **0.6325** | 0.6128 | 0.5965 | 0.5640 | 0.5838 | 0.6261 |
| Recall_p | 0.7713 | **0.8295** | 0.7558 | 0.7868 | 0.8023 | **0.8547** | 0.8295 | 0.6609 |
| F1 score_p | **0.7001** | 0.6900 | **0.6878** | 0.6878 | 0.6823 | 0.6792 | 0.6850 | 0.6425 |
| | Class Distribution: 10:90 split | | | | | | | |
| Measure | Ensemble CNN | Ensemble ML | char_aux | char_cnn | word_aux | SVM | Random Forest | Naive Bayes |
| Accuracy | **0.8728** | 0.8636 | 0.8638 | 0.8664 | 0.8445 | 0.8355 | 0.8592 | **0.8961** |
| Precision_p | **0.4338** | 0.3975 | 0.4112 | 0.4153 | 0.3760 | 0.3609 | 0.3875 | **0.4762** |
| Recall_p | **0.7281** | 0.6754 | 0.7368 | 0.7346 | 0.7171 | **0.8114** | 0.6776 | 0.2939 |
| F1_score_p | **0.5389** | 0.4999 | 0.5243 | **0.5275** | 0.4882 | 0.4990 | 0.4925 | 0.3611 |

The Krippendoff's Alpha is sensitive to data imbalance; and **(3)** We focus on sparse and imbalanced data distributions. For the 1,000 tweets for quality check, we got the Kappa score of 0.910 between our final labels and the labels we obtain from AMT. This is showing that our annotation guide was followed consistently by both our annotators and AMT Workers.

To simulate the data imbalance scenarios, we configured the class distribution and pre-sampled the dataset into six blocks for each distribution scenario, for 6-fold cross-validation. Each model was trained and tested on the same sets of training and test data to ensure a fair comparison. The number of data points included in each distribution scenario was maximized, but it was inevitably different between scenarios. Table 2 shows the dataset in each class distribution scenario.

*Drug Abuse Detection Results*

Table 3 shows the results for all individual models and two ensemble models. The ensemble model results are separated from the individual models for easier viewing. The highest value of each measure is marked in bold font. There is an interesting trend in the results of ensemble models. When the data is balanced or nearly balanced, the traditional ensemble machine learning model has a better performance than the ensemble deep learning model. At 50:50 and 40:60 splits, the ensemble machine learning model is superior over the ensemble deep learning model for most of the criteria. This is partially due to the relatively small dataset size. When the data becomes more imbalanced, e.g., at a 30:70 split, the ensemble deep learning model becomes better and has a higher F1-score for positive labels, compared with the traditional ensemble machine learning model. At 20:80 and 10:90 splits, the ensemble deep learning model takes the lead, most significantly in each measure for positive labels. The larger model capacity and the ability of the deep learning models to learn more complex non-linear functions can better distinguish the semantic differences between positive tweets and negative tweets, when the distribution of classes is heavily imbalanced.

Looking at individual machine learning models, Random Forest and SVM show a strong performance on all datasets, and they are especially good when the dataset is balanced. Naïve Bayes also has a good performance on a balanced dataset, but on an imbalanced dataset, it is heavily biased towards negative labels and has a poor performance for positive labels. Deep learning models generally have more stable performance, compared to traditional machine learning models, across all datasets, and a smaller difference between precision and recall, but their peak performances are not as good. Comparing between deep learning models, auxiliary features do not give C-CNN significant performance boost, and W-CNN is also not as good as the C-CNN model. However, in additional results that are not shown in this paper due to space limitations, auxiliary features give the plain W-CNN model a performance boost.

By investigating the performance of each individual model and the ensemble model that includes it, we can see that our ensemble strategy works well for deep learning models, as most of the measures for the ensemble model are higher than for any of its components corresponding measures. This effect was only observed a few times for traditional machine learning models. We expect that, by using more complicated ensemble strategies, deep learning has the potential to reach an even better performance level.

## Conclusions

In this study, we investigated how the data imbalance issue influences the performance of classifiers that are trained for identifying tweets that are related to drug abuse. We first collected a dataset with a broad selection of drug abuse-related keywords and slang terms. We explored the use of the Amazon Mechanical Turk platform as a reliable source for acquiring human-labeled tweets, and we obtained a solid dataset. We designed an ensemble deep learning classification model with both word-level and char-level CNNs, and we conducted a direct comparison with traditional machine learning models on our dataset, with simulated class imbalance. Experimental results show that our ensemble deep learning models have better performance than traditional machine learning models when the data is off-balance. Results also show that the ensemble strategy we used is effective for improving deep learning models. Finally, our analysis of the collected three million tweets, labeled by our model, shows an interesting temporal pattern that agrees with our intuition.

## Acknowledgements

The authors gratefully acknowledge the support from the National Science Foundation (NSF) grants CNS-1650587, CNS-1747798, CNS-1624503, and CNS-1850094.

## References

[1] H.P. Office, HHS Acting Secretary Declares Public Health Emergency to Address National Opioid Crisis, (2017). https://www.hhs.gov/about/news/2017/10/26/hhs-acting-secretary-declares-public-health-emergency-address-national-opioid-crisis.html.

[2] T.S. Abuse, and M.H.D. Archive, National Survey on Drug Use and Health, (2018). https://www.samhsa.gov/data/data-we-collect/nsduh-national-survey-drug-use-and-health.

[3] N.I. on Drug Abuse, Overdose Death Rates, (2018). https://www.drugabuse.gov/related-topics/trends-statistics/overdose-death-rates.

[4] G.V. Archive, 2015 Gun Violence Archive, (2018). http://www.gunviolencearchive.org/past-tolls.

[5] J.K. O'Donnell, J. Halpin, C.L. Mattson, B.A. Goldberger, and R.M. Gladden, Deaths Involving Fentanyl, Fentanyl Analogs, and U-47700-10 States, July-December 2016., *MMWR Morb Mortal Wkly Rep.* **66** (2017) 1197–1202.

[6] B. Hansen, K.S. Miller, and C. Weber, Early Evidence on Recreational Marijuana Legalization and Traffic Fatalities, Nat'l Bu. of Econ. Res., 2018.

[7] E.J. D'Amico, A. Rodriguez, J.S. Tucker, E.R. Pedersen, and R.A. Shih, Planting the seed for marijuana use: Changes in exposure to medical marijuana advertising and subsequent adolescent marijuana use, cognitions, and consequences over seven years, *Drug Alcohol Depend.* **188** (2018) 385–391.

[8] T. Sakaki, M. Okazaki, and Y. Matsuo, Earthquake Shakes Twitter Users: Real-time Event Detection by Social Sensors, in: Proc. 19th Int. Conf. WWW, Raleigh, North Carolina, USA, 2010: pp. 851–860.

[9] A. Sarker, and others, Social media mining for toxicovigilance: automatic monitoring of prescription medication abuse from Twitter, *Drug Saf.* **39** (2016) 231–240.

[10] S. Aslam, Twitter by the Numbers, (2018). https://www.omnicoreagency.com/twitter-statistics/.

[11] A. Arbor, Monitoring The Future, (2018). http://www.monitoringthefuture.org/.

[12] U. FDA, MedWatch: The FDA Safety Information and Adverse Event Reporting Program, (2018). https://www.fda.gov/safety/medwatch/.

[13] D.S. Hasin, and others, US adult illicit cannabis use, cannabis use disorder, and medical marijuana laws: 1991-1992 to 2012-2013, *Jama Psychiatry*. **74** (2017) 579–588.

[14] P. Seth, R.A. Rudd, R.K. Noonan, and T.M. Haegerich, Quantifying the Epidemic of Prescription Opioid Overdose Deaths, *Am J Public Health Res.* **108** (2018) 500–502.

[15] M. Chary, N. Genes, C. Giraud-Carrier, C.L. Hanson, L.S. Nelson, and A.F. Manini, Epidemiology from Tweets: Estimating Misuse of Prescription Opioids in the USA from Social Media, *J Med Toxicol*. **13** (2017) 278–286.

[16] C.L. Hanson, S.H. Burton, C. Giraud-Carrier, J.H. West, M.D. Barnes, and B. Hansen, Tweaking and tweeting: exploring Twitter for nonmedical use of a psychostimulant drug (Adderall) among college students, *J Med Internet Res*. **15** (2013).

[17] L. Shutler, L.S. Nelson, I. Portelli, C. Blachford, and J. Perrone, Drug use in the Twittersphere: a qualitative contextual analysis of tweets about prescription drugs, *J Addict Dis*. **34** (2015) 303–310.

[18] D. Mahata, J. Friedrichs, Hitkul, and R.R. Shah, #phramacovigilance -Exploring Deep Learning Techniques for Identifying Mentions of Medication Intake from Twitter, *CoRR*. (2018). http://arxiv.org/abs/1805.06375.

[19] H. Hu, and others, Deep Learning Model for Classifying Drug Abuse Risk Behavior in Tweets, in: 2018 IEEE Int. Conf. Healthcare Inform. (ICHI), IEEE, 2018: pp. 386–387.

[20] T. Katsuki, T.K. Mackey, and R. Cuomo, Establishing a link between prescription drug abuse and illicit online pharmacies: analysis of Twitter data, *J Med Internet Res*. **17** (2015).

[21] O. of the Law Revision Counsel, Drug Abuse Prevention and Control. Definitions. 21 U.S.C Sect. 802, (2018).

[22] M. Buhrmester, T. Kwang, and S.D. Gosling, Amazon's Mechanical Turk: A New Source of Inexpensive, Yet High-Quality, Data? *Perspect Psychol Sci*. **6** (2011) 3–5.

[23] X. Zhang, J.J. Zhao, and Y. LeCun, Character-level Convolutional Networks for Text Classification, *CoRR*. (2015). http://arxiv.org/abs/1509.01626.

[24] NoSlang.com, Drug Slang Translator, (2018). https://www.noslang.com/drugs/dictionary.php.

[25] G.A. Miller, WordNet: a lexical database for English, *Commun ACM*. **38** (1995) 39–41.

[26] Y. Fan, Y. Zhang, Y. Ye, and X. Li, Automatic Opioid User Detection from Twitter: Transductive Ensemble Built on Different Meta-graph Based Similarities over Heterogeneous Information Network., in: IJCAI, 2018: pp. 3357–3363.

[27] L.A. Jeni, J.F. Cohn, and F. De La Torre, Facing Imbalanced Data–Recommendations for the Use of Performance Metrics, in: 2013 Humaine Association Conference on Affective Computing and Intelligent Interaction, IEEE, 2013: pp. 245–251.

[28] K.A. Hallgren, Computing inter-rater reliability for observational data: an overview and tutorial, Tutorials in Quantitative Methods for Psychology. 8 (2012) 23.

[29] X. Zhang, J.J. Zhao, and Y. LeCun, Character-level Convolutional Networks for Text Classification, CoRR. (2015). http://arxiv.org/abs/1509.01626.

**Address for correspondence**

Corresponding author: NhatHai Phan, New Jersey Institute of Technology, NJ, USA, Email: phan@njit.edu